# Viscous Flow in Glass-Forming Liquids: The Twice-Activation Analysis and the Bond Wave Model

Elena A. Chechetkina

*Institute of General and Inorganic Chemisctry of Russian Academy of Sciences (1980-2011)*

Moscow, Russia

eche2010@yandex.ru

*Abstract* – **The "twice-activation" method for analysis of experimental viscosity-temperature data reveals a set of interconnected parameters describing the state of the flowing glass-forming liquid in terms of convergation point which describes an infinite set of $\eta(T)_i$ relations for the liquid considered. The observed uncertainty in the viscosity-temperature behavior permits to consider glass-forming liquid as the self-organizing system realizing by the bonds wave as dissipative pattern. The acoustic bond wave and the switching bond wave are considered generally and in different groups of glass-formers: inorganic, organic and polymers. The demonstrated correlation between two coordinates of convergation point and "kinetic" and "thermodynamic" measures of fragility permits to resolve the problem of the measures' discrepancy.**

## 1. INTRODUCTION

There is a wide-accepted tendency to consider glass as an intermediate state between crystal and liquid, a state that represents either a *non-crystalline solid* or as a *frozen liquid*. Note, however, that glass-forming liquids is an unusual liquid that demonstrates an extremely high viscosity even above melting point and the non-Arrhenius viscosity-temperature dependence. Both features attract a continuous interest of glass scientists [1-5], being an unquenchable source for numerous theories/models of glass formation and glass transition, a situation indicates a poor understanding of glass nature in frames of the existing "classical" approaches.

The non-classical approach used here is based on the specificity of chemical bonding in glass-forming substances, combined with the notions about self-organization. The bond specificity represents as the two-state bonding, when basic bond, the same that in related crystal, can transform reversibly into an alternative bond. This specificity was investigated by *Dembovsky* who have intensively studied hypervalent configurations as alternative bonds in inorganic glass-formers [6-8]. My contribution is the self-organization of alternative bonds in the form of the bond wave as characteristic dissipative pattern – the reader can find a brief description and application of the bond wave model in [9]. Here the model is developed for the case of viscous flow.

The paper is organized as follows. In **Section 2** the *twice-activation* (TA) method is demonstrated on the example of "strong" $GeO_2$ and "fragile" glycerol. Besides the difference in fragility and chemical bonding, both liquids obey to a standard behavior, giving a set of lines {$A_i;G_i$} in the TA-plot and then the master line in semi-logarithmic log$A=f(G)$ coordinates. The line parameters are considered as coordinates of *convergation point* {$a;b$}, and the point divergence gives the *viscous space* for the liquid. The set of 11 typical glass-formers, from "strong" $SiO_2$ to "fragile" ortho-terphenyl, presented in *convergation plot* reveals a sharp difference between inorganic and organic groups beyond the difference in fragility.

In **Section 3** the formal results are interpreted by means of the bond wave model. In terms of self-organization, convergation point {$a;b$} represents the *attractor* for the process of viscous flow, and $A_i$ and $G_i$ represent the *order parameter* and the *managing parameter* for the {$A_i;G_i$} viscous pattern. Two types of bond waves, the *acoustic* bond wave and the *switching* bond wave, are considered in accord with the chemical bonding peculiarities in different groups of glass-formers. In **Section 4** the results are discussed from the fragility point of view.

## 2. THE TWICE-ACTIVATION ANALYSIS

### 2.1. Twice-Activation Plot

The *Arrhenius* character of viscous flow implies fulfilment of the Arrhenius equation

$$\eta(T) = \eta_{Arr} \cdot \exp(E_{Arr}/RT) \quad (1)$$

with constant pre-exponent $\eta_{Arr}$ and activation energy $E_{Arr}$.

This equation is unsuitable for glass-forming liquids, which demonstrates deviations from a line in Arrhenius coordinates log$\eta=f(1/T)$ in a wide enough temperature interval since both "effective" activation energy $E_\eta$ defining as the slope of experimental curve and related pre-exponent are temperature dependent. Therefore, the analysis of temperature dependence for only $E_\eta$ (e.g. [10]) is only a part of the $\eta(T)$ analysis. In order to avoid two-factor analysis of $\eta(T)$, I have proposed earlier [11] to fix pre-exponent by means of the *Eyring* equation [12]

$$\eta(T) = \eta_E \cdot \exp(E_E/RT) \quad (2),$$

where $\eta_E = Nh/V$ ($N$ and $h$ are the Avogadro's and Plank's constants, and $V$ is the molar volume) is practically constant because of a negligible temperature dependence of density $\rho \sim 1/V$ as compared with the viscosity temperature dependence. Using calculation formulae

$$\eta_E \text{ [poise]} = 0.0039 \rho \text{ [g/cm}^3\text{]} / M \text{ [g]} \quad (3)$$

one obtains $\log \eta_E$(poise)= −3.9 for $GeO_2$, −3.65 for Se; −4.25 for glycerol. Noticeably, $\log \eta = -4$ corresponds to the infinite-temperature limit for viscosity in the Angell plot [1].

The second step of analysis is substitution of experimental $\eta(T)$ points by the $E_E(T)$ points calculated as

$$E_E \text{ [kcal/mole]} = 4{,}573 \cdot T \, [\log \eta(T) - \log \eta_E]/1000 \quad (4).$$

This step would look as a meaningless procedure unless the fact that the $E_E(T)$ points arrange in a line in the "twice activation" plot (*activation* plot for *activation* energy), as it is shown in **Fig.1** on the example of glycerol.

One can say that the $\log \eta (1/T)$ dependences in **Fig.1a** looks linear too. Note, however, that every liquid can demonstrate the Arrhenius-like behavior in a less or more wide temperature interval depending on extent of fragility. As to the considered glycerol, its non-Arrhenius behavior becomes clear when going deeper into the high-temperature region – see **Fig.16** below, from which only the first point of the 0.0 GPa curve is presented in **Fig.1a**.

The third step reveals the activation-type relation between all the obtained TA-lines:

$$E_E(T)_i = A_i \cdot \exp(G_i/RT) \quad (5)$$

or **master line** in semi-logarithmic coordinates:

$$\log A = a - b \cdot G \quad (6).$$

The master lines for glycerol and $GeO_2$ are shown in **Fig.2**. The lines parameters, {0.873; 0.779} and {1.775; 0.137} represents coordinates of **convergation point** {$a;b$}.

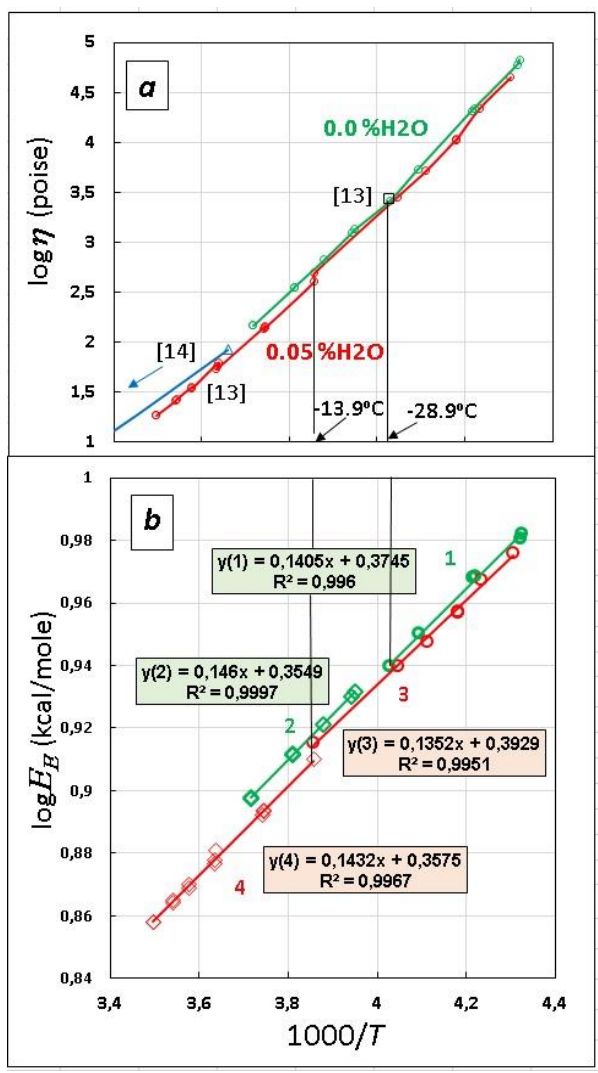

**Figure 1.** The viscosity-temperature data for glycerol after [13,14] presented in the *Arrhenius* plot (*a*) and TA-plot (*b*). The lines' are drawn and characterized by *Excell* program.

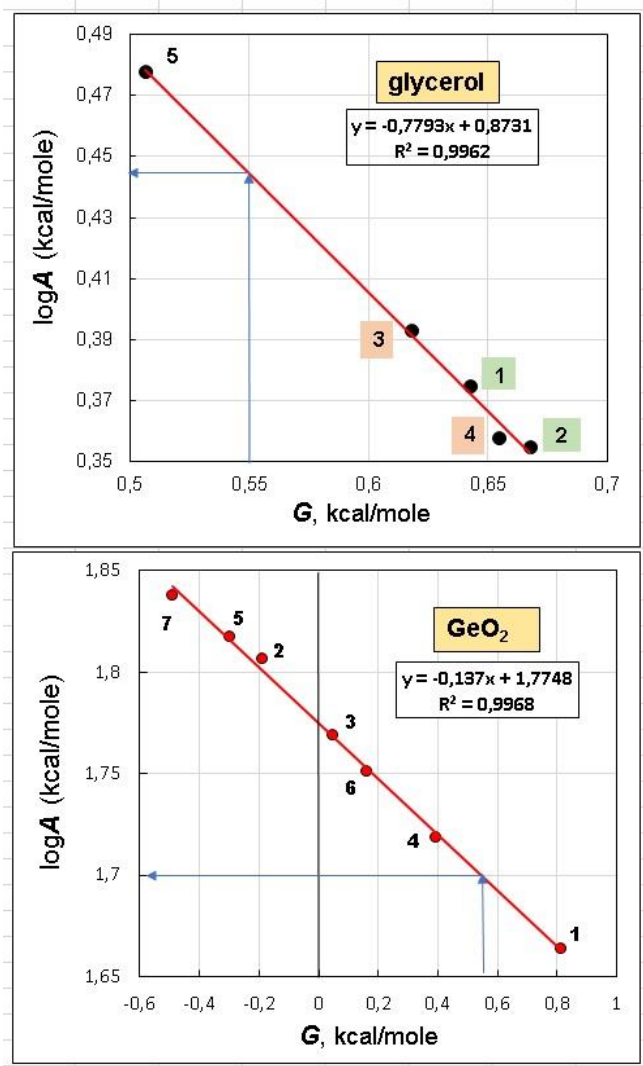

**Figure 2.** Master plots for glycerol (points 1-4 after **Fig.1**, point 5 after digital data from [14]) and $GeO_2$ (the figure is reproduced from [9]).

## 2.2. Viscous Space

As well as experimental viscosity-temperature curves are merged into convergation point, the point can diverge into an infinite set of the curves described by a general equation

$$\log\eta(T)_i = \log\eta_E + (A_i/2.303\,RT)\cdot\exp\{[(a-\log A_i)/b]/RT\}$$
$$i=1,2,\ldots\infty \quad (7)$$

which has two variables: temperature and energy $A_i$ of the $i$-th TA-line corresponding to Eq.(5). Formally, this is the three-parameter equation ($\eta_E$, $\{a;b\}$, $A_i$) like classical VFT equation, $\log\eta=\log\eta_\infty+A/(T-T_0)$, or the equations proposed by *Avramov & Milchev* [3] or *Mauro et al* [4]. The principal difference is a relatively free character of Eq.(7): despite of a fixed convergation point for the substance, one cannot predict what $A_i$ will realize in a concrete experiment.

The freedom of realization, however, is restricted by some reasons since the experimentally observed curves, $\eta^{exp}(T)$, tends to group around the most probable curve considered as a "true" viscosity-temperature dependence. Let us consider the location of experimental curve(s) in viscous space by means of **Table 1** and **Fig.3** using the same exemplar pair of "fragile" glycerol and "strong" $GeO_2$, the same $G_i$ interval (from −1 kcal/mole to +1 kcal/mole in concord with **Fig.2**) and the reduced temperature $T^*$ corresponding to equal viscosity $\log\eta(T^*)$ for calculated curves, as it is shown in **Fig.3**. The reduced temperature is determined by $b$-coordinate of convergation point by equation

$$T^* = (1/2.303\,R)/b = 218.7/b \quad (8)$$

which follows from Eqs.(2,5,6), $R=1.9858\cdot10^{-3}$ [kcal/mole·K] and $\ln 10=2.303$.

**Table 1**. Divergence of viscosity for glycerol ($T^*$=280K) and $GeO_2$ ($T^*$=1596K). $G$ and $A$ in [kcal/mole], $\eta$ in [poise].

|  | $\log\eta(T)_i$ after Eq.(7) for **glycerol** | | |
|---|---|---|---|
|  | $G = -1$ | $G = 0$ | $G = +1$ |
|  | $A = 45$ | $A^0 = 7.4$ | $A = 1.2$ |
| 0.9$T^*$ | 1.05 | 2.20 | 3.60 |
| $T^*$ | 1.57 | 1.57 | 1.57 |
| 1.1$T^*$ | 1.98 | 1.03 | 0.23 |
|  | $\log\eta(T)_i$ after Eq.(9) for **$GeO_2$** | | |
|  | $G = -1$ | $G = 0$ | $G = +1$ |
|  | $A = 82$ | $A^0 = 60$ | $A = 44$ |
| 0.9$T^*$ | 4.86 | 5.17 | 5.50 |
| $T^*$ | 4.26 | 4.26 | 4.26 |
| 1.1$T^*$ | 3.73 | 3.52 | 3.31 |

In **Table 1** one can see the first difference between the liquids considered: for the same $\Delta G$ and $\Delta T^*$ intervals there is the 38 times change of $A$ for "fragile" glycerol (from 1.2 to 45 kcal/mole) and only the 2 times change for "strong" $GeO_2$ (from 44 to 82 kcal/mole). The change in viscosity is in 4.6 times and 1.5 times respectively.

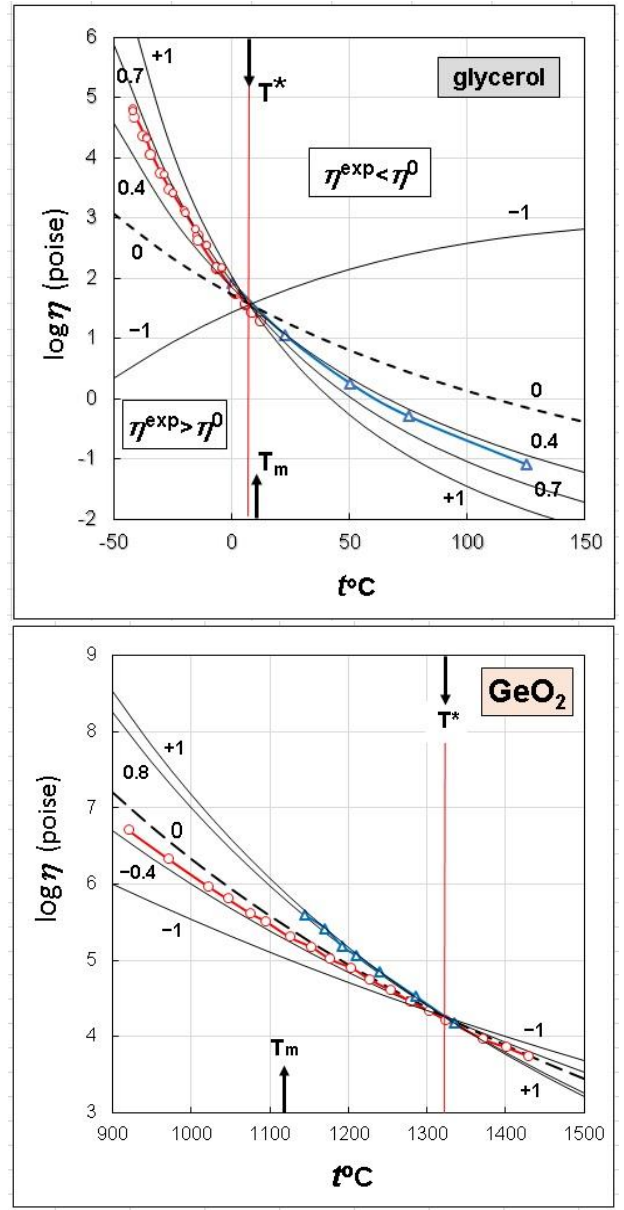

**Figure 3**. Viscous space for glycerol ($\log\eta_E$=−4.25 and {0.873; 0.779}) and $GeO_2$ ($\log\eta_E$=−3.9 and {1.775; 0.137}). Experimental $\eta(T)$ points are after *Tammann & Hesse* [13] (red circles and curve) and *Cook et al* [14] (blue triangles and curve) for glycerol, and after *Bruckner* [15] (red points) and *deNeufville et al* [16] (blue triangles) for $GeO_2$.

By means of **Fig.3** one can note the second difference: the viscous space for "fragile" glycerol is much wider than for "strong" $GeO_2$. Although both liquids tend to arrange near the pointed *standard curve* of $G$=0 which corresponds to equation

$$\eta^0(T) = \eta_E\cdot\exp(A^0/RT) \quad (9),$$

the deviations of $\eta^{exp}(T)$ from $\eta^0(T)$ when removing from $T^*$ grows much faster for "fragile" glycerol.

## 2.3. Convergation Plot

Convergation points calculated for a set of typical glass-formers are given in **Table 2** together with general temperature points: glass-transition temperature $T_g$, melting point $T_m$, and temperature of equal viscosity $T^*$ by Eq.(8).

**Table 2**. Convergation points for typical IFG and OGF (Gly is glycerol, DBP is di-n-butylphthalate, OTP ortho-terphenyl)

| Subst. | $T_g$, K | $T_m$, K | $T^*$, K Eq.(8) | $a; b$ | Ref.for $\{a;b\}$ |
|---|---|---|---|---|---|
| $SiO_2$ | 1500 | 1983 | 2165 | 2.014; 0.101 | 9* |
| $GeO_2$ | 818 | 1389 | 1596 | 1.775; 0.137 | 9* |
| $BeF_2$ | 592 | 827 | 959 | 1.578; 0.228 | 17,18 |
| $B_2O_3$ | 550 | 748 | 734 | 1.486; 0.298 | 9* |
| $As_2S_3$ | 460 | 572 | 524 | 1.511; 0.417 | 19 |
| $As_2Se_3$ | 450 | 645 | 511 | 1.451; 0.428 | 20,21 |
| Se | 310 | 494 | 312 | 1.324; 0.700 | 9* |
| Gly | 190 | 291 | 280 | 0.873; 0.779 | 13,14 |
| Salol | 210 | 310 | 300 | 0.728; 0.749 | 22-24 |
| DnBP | 170 | 238 | 251 | 0.760; 0.873 | 25,26 |
| OTP | 240 | 331 | 332 | 0.789; 0.659 | 23,24 |

* The referencies for $\eta(T)$ data see in [9].

The substances in Table 2 arrange with increasing fragility, and one can see the opposite tendency for $a$ and $b$ values when passing "strongest" $SiO_2$ to "fragile" o-terphenyl (OTP). A more detailed presentation in **Fig.4** reveals an obvious distinction between inorganic and organic glass-formers: while the formers occupy systematically "strong" and intermediate regions of convergation plot, the latters dispose chaotically in the "fragile" region.

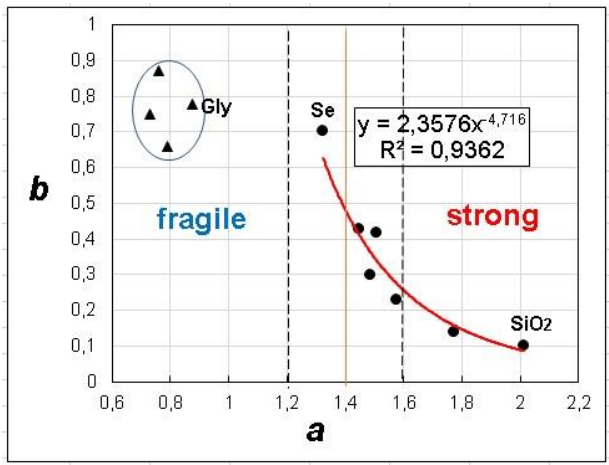

**Figure 4.** Convergation plot for the substances from **Table 2**.

# 3. THE MODEL FOR VISCOUS FLOW

## 3.1. Convergation Point and Self-Organization

In terms of synergetics, a general theory for self-organization [27,28], convergation point represents an **attractor** for the viscosity-temperature behavior. This attractor generates an infinite set of *viscous patterns* $\eta(T)_i$ defined by $A_i$ in Eq.(7) which therefore can be considered as the **order parameter**, whose value relates to the viscous pattern energy.

The energy varies in accord with the master line of a substance (e.g. from 46.1 to 54.6 kcal/mole for $GeO_2$ and from 2.26 to 2.98 kcal/mole for glycerol in **Fig.2**) depending on the $G$ value, whose variation is much lower, from −1 kcal/mole to +1 kcal/mole for the considered pair. Therefore, $G$ can be considered as the **managing parameter** for the process of viscous flow, whose value defines the viscous pattern energy $A_i$ (see blue thin arrows in **Fig.2**) and whose sign indicate the flow mechanism. To understand the mechanism one should propose an adequate model of viscous flow, a model that describes both viscous pattern, including different patterns realizing in organic and inorganic glass-formers, and mechanism of viscous flow that correspond negative or positive sign of the order parameter.

## 3.2. Acoustic Bond Wave

The well-known examples of self-organizing patterns are the *cellular structures* for the *Bernard* convection and *chemical waves* realized by the *Belousov-Zhavitincky* reactions [28-30]. Earlier I have proposed the **bond wave** as the dissipative pattern characteristic for glass-forming substances [31, 9]. Bond wave means the spatiotemporal correlation between elementary acts of bond exchange, $\Sigma\Sigma(BB\leftrightarrow AB)$, where BB is the basic bond, the same that in related crystal, and AB is *alternative bond* that exists only in glass-forming liquid or glass. This is the basic model which should be adapted to the process of viscous flow, beginning from the elementary act of bond exchange.

The basic bond for *inorganic glass formers* (**IGF**) is usual two-center two-electron (2c-2e) covalent bond, CB. The higher-energy alternative bond in IFG *hypervalent bond*, HVB, after *Dembovsky* [6-8]. Then the bond wave in IGF may be denoted as $\Sigma\Sigma(CB\leftrightarrow HVB)$. A simple model for elementary act of bond exchange is shown in **Fig.5** using three-center four-electron (3c-4e) bond as HVB.

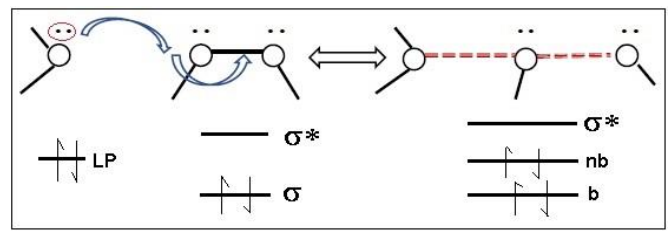

**Figure 5**. Elementary act of bond exchange in Se after [6]; LP is the lone-pair atomic orbital, *nb* and *b* are non-bonding and bonding orbitals for TCB.

Another situation is in *organic glass formers* (**OGF**), whose solid or liquid state is provided by intermolecular bonds, which are strong enough to fix structure at a relatively high temperature $T>200$-$300$K. Two types of intermolecular bonds in OGF are illustrated by **Fig.6**: hydrogen bond (H-bond) due to proton delocalization between two electronegative atoms (two oxygen atoms in the case of water or alcohol) and $\pi$-$\pi$ bond due to interaction between adjacent aromatic rings.

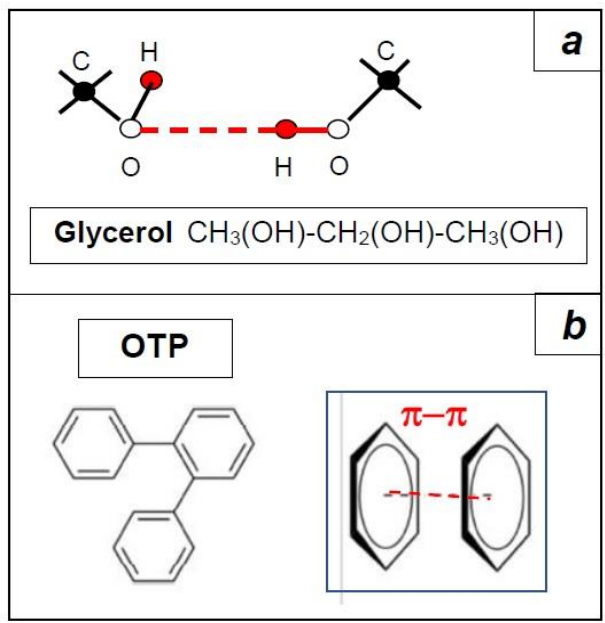

**Figure 6.** Intermolecular bonding (red pointed lines) realized by H-bond or $\pi$-$\pi$ stacking in glycerol (*a*) or *o*-terphenyl (*b*).

Both types of intermolecular bonding can coexist in OFG – for the substances from **Table 2** the mixed case of intermolecular bonding realizes in salol ($C_{13}H_{10}O_3$, aromatic alcohol) and DnBP (di-*n*-butyl-phthalate $C_{16}H_{22}O_4$, ester).

The nature of alternative bonds in OGF, which are the necessary condition of the bond wave, is under the question. By analogy with IGF, the high-viscous organic liquid can keep a continuous network by means intermediate complexes of intermolecular H-bonds and/or $\pi$-$\pi$ bonds. These complexes represent *alternative bonds* for OGF The reader can find them in a vast massive of literature dedicated to chemical bonding in water and aromatics (see, e.g., [32-34]).

The situation for *glass-forming polymers* containing long replicated chains is specific. A customary approach is to relate high viscosity with a high dimension of polymeric "molecules", particularly, with long chains which are therefore entangled giving an "amorphous" state. Such a mechanistic approach was applied even for Se considered as "inorganic polymer" [35]. The analogy, however, fails when passing to other inorganic glass-formers having no the chain-type crystalline counterpart. Nevertheless, the notions about a "polymeric" network of inorganic glass-forming liquids remains popular up to now, a situation that prevents recognizing of the groups' specificity.

As to the glass forming polymers one should search alternative bonding states arising between extra-large molecules in local and collective way, the latter in the form of the bond wave. These states may use the H-bond and/or $\pi$-bond components as the triggers for local and collective structure rebuilding. The example of $\pi$-trigger is polysterene, a classical glass former whose chains contain lateral aromatic rings arranged randomly (amorphous "atactic" phase [36], $T_g$=100ºC) or ordinarily (isostatic and syndiotactic crystalline phases, $T_m$=240 and 270ºC) when the rings are ordered in one or another way along the [-$CH(C_6H_5)$-$CH_2$-]$_n$ chains.

Despite of the bonding specificity, one can describe all groups of glass formers in a general way when using the bond wave model. As far as concentration of alternative bonds (AB) increases with temperature, the higher temperature the larger space where AB can act collectively. Namely, above $T_g$ the 3D bond waves of the **Λ3**(*T*) wavelength (**Fig.7a**) animates all the structure, and below $T_g$ the distance between the layer-like wavefronts/layers become so large that they cannot "feel" each other, and the 3D wave stops its propagation.

Glass transition does not forbid structure mobility completely since there remains 2D bond waves of the **Λ2**(*T*) wavelength that spreads along the layers with the V2(**T**) velocity even at $T<T_g$ (**Fig.7b**). The 2D bond waves also freeze below **T2**<**T_g**, and so on up to very-low temperatures $T<$**T1** when even 1D bond waves, representing collective moving along the strings, freeze. This step-like process is described in more details in [9, 31].

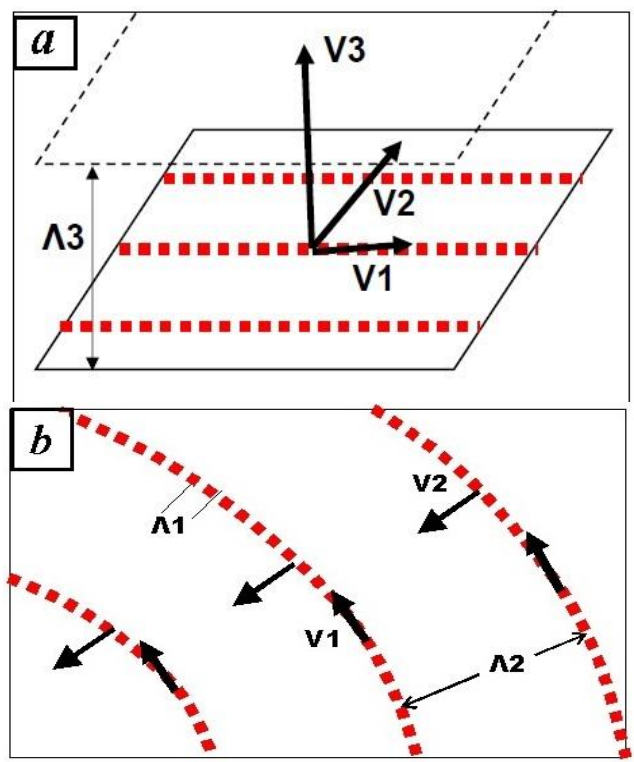

**Figure 7**. Bond waves of different dimensionality and related temperature-dependent parameters – velocity **V** and the wavelength **Λ**. Small red squares are alternative bonds.

One can name the above-described 3D bond wave the **acoustic bond wave** (ABW) by analogy with ordinary acoustic wave representing a system of alternating stretching-squeezing regions. Really, the wavefronts of 3D bond waves populated with alternative bonds can be associated with equidistant stretching regions – just because they are weaker and longer than basic bonds in the rest network (see, e.g., three-center bonds and covalent bond in **Fig.5**).

### 3.3. Switching Bond Wave

As far as alternative bonds are the high-energy states in the network, they are the active places for network rebuilding both local and collective. If the collective processes are provided by the bond waves, then *viscous flow* is stipulated by 3D bond wave animating all the volume at $T>T_g$, and *plastic flow* is the process below $T_g$ stipulated by 2D bond waves spreading along the stopped layers – the wavefronts of the frozen 3D bond wave – see **Fig.7b**. The problem is that the elementary act of ABW (e.g., that shown in Fig.5) is reversible, so it cannot ensure mass transport needed for flow. Therefore, to describe viscous flow one needs another bond wave – the **switching bond wave** (SBW) whose elementary act is irreversible switching of basic bond, for example from atoms 1-2 to 2-3 in **Fig.8**.

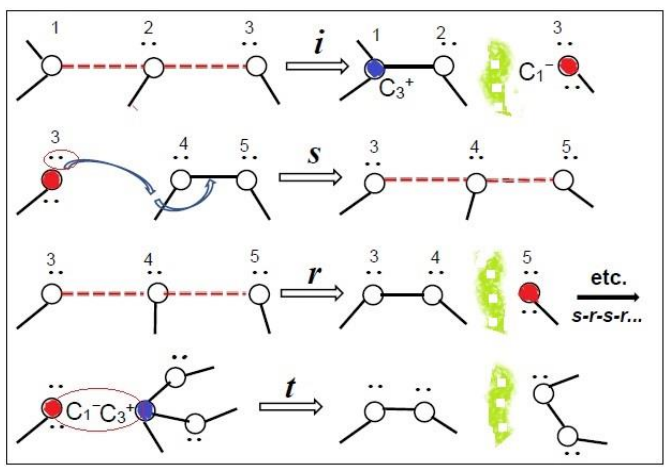

**Figure 8**. Elementary act of *bond switching* in Se using TCB as intermediate. The space between neighbor Se-chains is marked by the yellow-green snakes.

In this scheme, which was proposed firstly by *Dembovsky* [6] there is used also famous valence alternation pair (VAP) after *Kastner, Adler & Fritzsche* [37], where $C_1^-$ and $C_3^+$ are one- and three-coordinated atoms. The preceding state is alternative bond for ABW (see **Fig.5**). Irrespectively of the nature of alternative bond and intermediates, elementary act of bond switching includes four steps: *i* – initiation, *s* – stacking, *r* – release, *t* -termination of Se-chain.

It seems reasonable to divide the *energy transfer* realizing by ABW in the process of collective bond exchange, and the *mass transfer* realizing by SBW in the process of viscous flow. Both transfers need information in which the waves should run. The information aspect of self-organization [38, 39] is realized here by the notions about *information field* that fives the wave direction [9]. Two information fields which usually present in glass practice, the temperature gradient (**grad***T*) and the pressure gradient (**grad***P*) are shown in **Fig.9**. The picture illustrates, first, a possibility for a direct observation of the frozen wavefronts by means of *fractography* and, second, the *solitonic* behavior of the bond waves that can intersect each other without distortion [9].

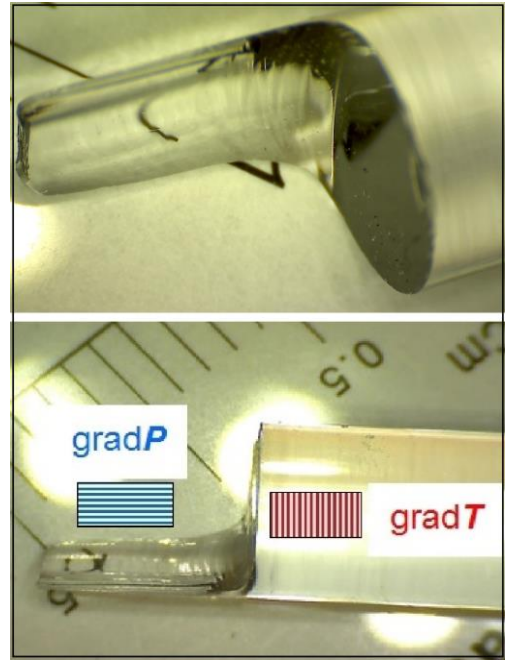

**Figure 9.** Typical fracture of glassy rod. See text for details.

### 3.4. The ABW/SBW Interaction

Although **Fig.9** shows different directions in which ABW and SBW run, the waves should be connected at least energetically, because just ABW transfers thermal energy for generation of alternative bond initiating the chain of switching (compare **Fig.5** and **Fig.8**). The scheme of ABW/SBW coupling during viscous flow is shown in **Fig.10** originated from **Fig.7**.

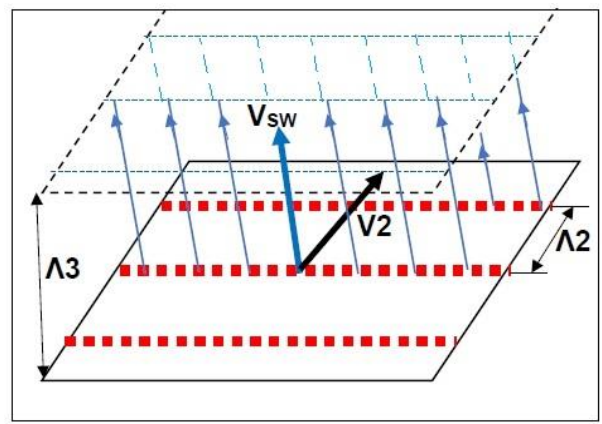

**Figure 10**. The scheme of coupling between ABW and SBW.

In **Fig.10** red points located on the 3D ABW wavefront are alternative bonds that give the switching chains shown by blue arrows in the **Vsw** direction. Each string on the ABW wavefront gives its own family of the switching chains, so connecting 2D ABW with the SBW wavelength as **Λ2=Λsw**.

Each switching chain shown by blue arrow in **Fig.10** represents an *autocatalytic chain reaction*, which is initiated in the CB→AB point/atom when the ABW wavefront passes through it. Then the AB energy in spent in the switching jumps by the switching agent ($C_1^-$ in **Fig.8**), whose concentration is negligible even when compared with the AB concentration. Therefore, the chain termination is caused not so by the poison ($C_1^-$ in **Fig.8**) but when the initial AB energy is exhausted. Then the *chain length L* depends on the AB energy and the energy needed for elementary jump, being *constant* for the substance considered. Then one can distinguish three regimes of viscous flow: the *waiting* regime when $L<Λ3$, the *acoustic* regime when $L=Λ3$ and the *overrunning* regime when $L>Λ3$, as it is illustrated in **Fig.11**.

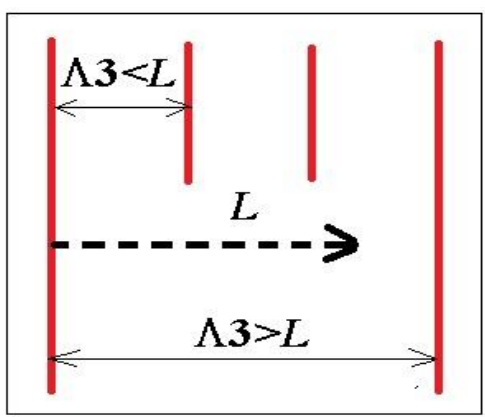

**Figure 11.** Three regimes of flow depending on the chain length *L* and the ABW wavelength **Λ3** relation.

At the *waiting* regime ($L<Λ3$) the family of chains exhausts its energy before coming of next ABW wavefront and waits when the wavefront bring a next portion of energy to continue switching. In the *overrunning* regime ($L>Λ3$) the ABW wavefronts continuously feed the chains which, therefore can develops infinitely – unless they meet a "poison" embedded in the network, e.g., $C_3^+$ for the switching chain shown in **Fig.8**. The boundary case of $L=Λ3$ call the *acoustic* regime because now SBW moves synchronically: the exhausted chain is fed simultaneously by the incoming ABW wavefront.

It seems reasonable that just *G*, the managing parameter, determines not only the energy of viscous pattern *A* by the *G* value by Eq.(6) but also the regime of flow by the *G* sign. Based on a general tendency of lowering *G* with temperature (a fine example is $B_2O_3$ in Fig.2 of Ref.11), one can suppose that the high-energy viscous patterns corresponding to the short-wavelength ABW [9] and so the $L>Λ3$ case realize in the overrunning $G<0$ region. Contrary, the low-energy viscous patterns realizing at $G>0$ correspond to the waiting regime.

### 3.5. Acoustic Flow and Adaptation Ability

The boundary case of $G=0$ is of a special interest. Corresponding "acoustic" viscous pattern of the $A^0=10^a$ energy describes by the Arrhenius-type Eq.(9). Using the *a* values presented in **Table 2** together with characteristic temperature, one obtains **Fig.12**, where an obvious difference between inorganic and organic glass-formers is seen as a strong dependence of $A^0$ on $T_m$ (and so on $T_g≈⅔T_m$) for IGF in contrast to indifference for OGF.

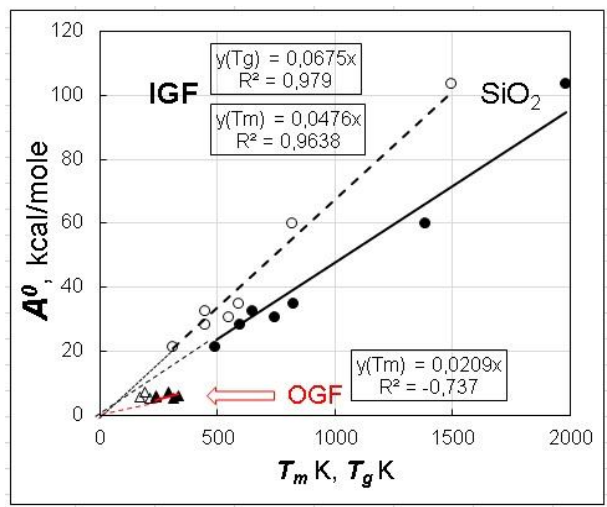

**Figure 12.** The acoustic pattern energy $A^0=10^a$ as a function of melting point and the glass transition temperatures for inorganic glass-formers (circles) and organic ones (triangles).

Interestingly, both $A^0(T_m)$ and $A^0(T_g)$ lines for IFG come into the zero point, so one can rewrite the $A^0(T_m)$ equation as $A^0=24RT_m$. The equation coefficient correlates with the well-known condition for glass formation $G',G''>30RT_m$ ($G'$ and $G''$ are the barriers for crystal nucleation and growth) after *Turnbull & Cohen* [40], and also to the $E_a>30RT_m$ condition ($E_a$ is atomization energy) which was obtained by me earlier for glass-forming oxides and elements (Se, S, P) [41].

As far as the acoustic bond wave is born in the subsystem of basic bonds (BB) which generates the alternative bonds subsystem, the IFG/OFG difference can be related with the basic bonds difference: BB=CB (covalent bonds) for IGF, and BB=IMB (intermolecular bonds) for OGF based on H-bonds and/or π-π interactions (see **Fig.6**). Really, the CB strength for IGF changes from 50 kcal/mole for Se-Se to 110 kcal/mole for Si-O correlates with $A^0$ from 20 kcal/mole to 120 kcal/mole for $SiO_2$, the deviation being increasing with fragility.

The strength of intermolecular bonds in OGF is lower and less differentiative. The H-bond energy depends strongly on the linking atoms, e.g., 39 kcal/mole for the F−H····F bond, 7 kcal/mole for O−H····N bond, 5 kcal/mole for O−H····O and 2 kcal/mole for N−H····O [42] it is about 5 kcal/mole for the OGF considered. Like H-bond, the **π-π** bond energy depends on the rest molecule and its surroundings; the **π-π** energy was estimated to vary from 0.5 to 13 kcal/mole in the case protein-porphyrin structures [43]. The $A^0$ energies for OGF in **Fig.12**, from 5.3 kcal/mole for salol to 7.5 kcal/mole for glycerol, correspond well to the above-estimated energies for the O−H⋯O bond and the $\pi$-$\pi$ bond of an average energy.

While $a=\log A^0$ defines the sense of *a*-coordinate of convergation point as the energy of "acoustic" viscous pattern, the sense of *b*-coordinate, which is expressed as

$$b = -(dA/dG) \quad (10)$$

defines the regulation strength of the managing parameter *G*. "Fragile" liquids, which can form viscous patterns in a wide interval of energy *A* for un a unit *G* interval possess a much higher *adaptation ability* to the flow conditions than "strong" liquids of a much less variation in the patterns' energy.

In contrast to a-coordinate of convergation point, b-coordinate of organic glass-formers depends on melting point too – see Fig.12 for OGF. The difference between inorganic and organic glass-formers, however, remains – not so much in position than in their different functional form, which is linear for OGF and the inverse degree for IGF. This means that coordinates of convergation point have not only a different sense in frames of the bond wave model but also reflect the different sides of fragility.

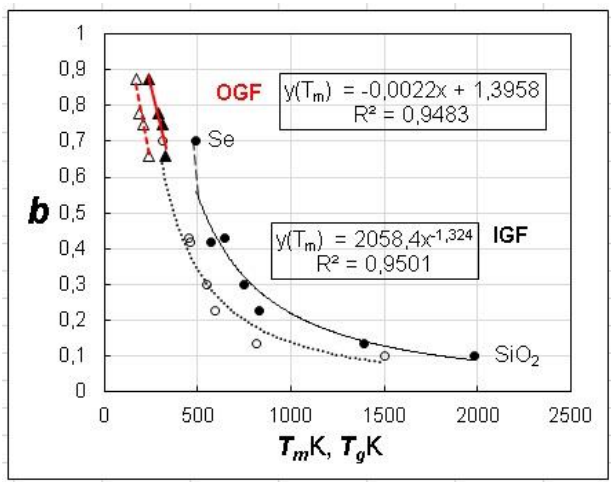

**Figure 13**. The *b*-coordinate of convergation point as a function of characteristic temperatures for organic (triangles) and inorganic (circles) glass formers from **Table 2**.

## 4. DISCUSSION
### 4.1. Convergation Point and Fragility

A popular concept of **fragility** is of the fifty years old [44] – and even older when one remembers *Oldenkop* who have used the "Angell" coordinates $\log\eta=f(T_g/T)$ as far as in 1957 for the analysis of silicate glass-formers [45]. Remember also that the terms "long" (strong) and "weak" (fragile) glasses are known far before in glass technology, and the fact that *Laughlin & Uhlmann* have used the same coordinates when considering viscosity of simple organic liquids [23]. It was *Angell* [1] however, who introduce fragility in glass science as a classification tool for glass-forming liquids of different nature – inorganic, organic (including polymers) and molecular (including water [46]. Now "fragility" becomes a widely-accepted tool for consideration of viscosity data and speculations about glass nature (e.g., [47-49]) in a wide class of materials including glass-forming metallic alloys [50]. Let continue this trend using convergation point and the bond wave model.

In **Table 3** the considered set of glass formers is characterized by specific viscosities and two basic measures/indexes of fragility, *m* and $\Delta C_{pg}$.

**Table 3.** Specific viscosities ($\eta_E$ is the *Eyring* pre-exponent, $\eta_m$ is viscosity at $T_m$, in [poises]) and two indices of fragility for typical inorganic and organic glass-formers..

| Subst. | $\log\eta_E$ | $\log\eta_m$ | *m** | *m* [51] | *m* [47] | $\Delta C_{pg}$ [52] |
|---|---|---|---|---|---|---|
| **SiO$_2$** | −3.8 | 7.8 | 19 | 20 | 25 | 0.144 |
| **GeO$_2$** | −3.9 | 5.5 | 18 | 20 | - | 0.50 |
| **BeF$_2$** | −3.8 | 6.8 | 22 | - | - | 0.00 |
| **B$_2$O$_3$** | −4.0 | 4.8 | 38 | 32 | 44 | 2.55 |
| **As$_2$S$_3$** | −4.3 | - | 29;31 | - | - | 3.1 |
| **As$_2$Se$_3$** | −4.3 | - | 38;40 | - | - | - |
| **Se** | −3.65 | 1.6 | 55 | 87 | 71 | 3.08 |
| **Gly** | −4.25 | 1.1 | - | 53 | 48 | 4.41 |
| **DnBP** | −4.8 | 1.5 | - | 69 | - | - |
| **Salol** | −4.6 | −1.0 | - | 73 | 66 | 3.31 |

* calculated from the same $\eta(T)$ data that were used for evaluation of convergation point {*a*;*b*} in **Table 2.**

First measure called usually *kinetic fragility* and defined as

$$m = \partial \log\eta(T)/\partial(T_g/T)|_{T_g} \quad (11)$$

represents simply the slope of the viscosity-temperature curve presented in the "*Angell plot*" at the $T_g/T=1$ point, i.e. at the glass transition temperature; this temperature corresponds to $\log\eta=13$ when $\eta$ in [poise] or $\log\eta=12$ for [Pa·s].

The second measure known as *thermodynamic fragility* is defined by the jump of heat capacity at glass transition. One should take in mind the regime of overcoming the glass transition region (cooling or heating) and a rate of the process. Note also that the jump may expressed in a relative $C_l/C_g$ (liquid/glass) or in absolute $\Delta C_{pg}$ units.

It is seen in **Table 3** that "strong" liquids (SiO$_2$, GeO$_2$, BeF$_2$) demonstrates substantially lower values of *m* and $\Delta C_{pg}$ than "fragile" liquids (from Se to OTP). This is an expected behavior which was emphasized firstly in 1976 in the classical work of *Angell & Sichina* – see Fig.6 in [53], which were reproducing later variations by others up to nowadays (see, e.g., [54]).

This relation, however, was found to be incorrect not only for different groups of glass-formers [55] but also in the same glass-forming system [56]. The uncertainty reveals also in our set of glass-formers presented in **Fig.14**, from which one can see both qualitative relation and quantitative unrelation for the considered set of the substances.

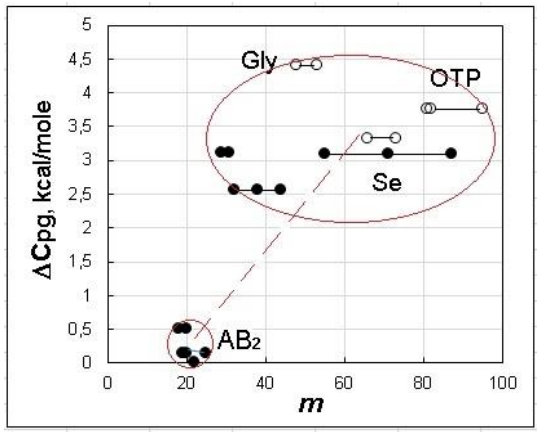

**Figure 14.** "Thermodynamic" fragility $\Delta C_{pg}$ versus "kinetic" fragility $m$ after **Table 3**.

There are two distant regions for the substance location: the narrow region in the form of small circle for typical "strong" liquids, and a wide ellipse for others. Although both inorganic and organic glass-formers can be presented together, the correlation between fragilities is bad.

The situation becomes better when the fragility indices are compared in pairs with coordinates of convergation point {$a;b$} as it is done in **Fig.15**.

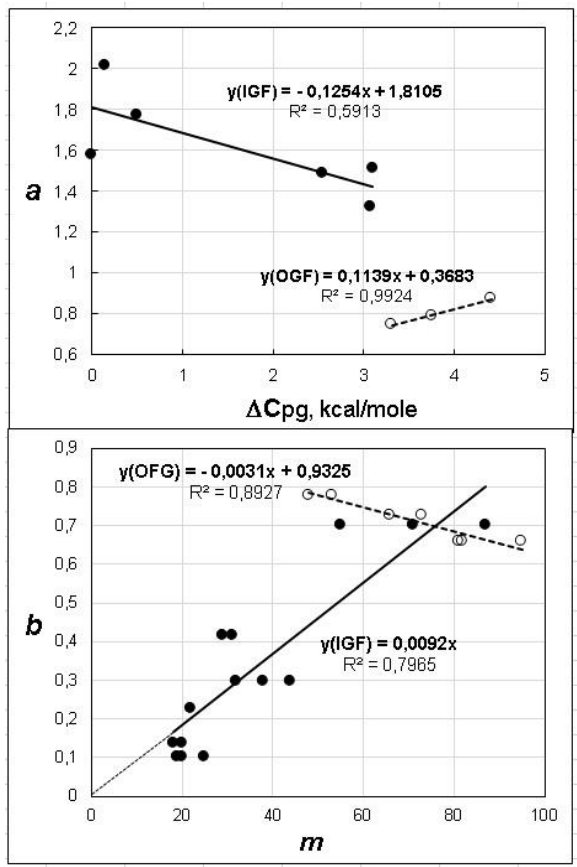

**Figure 15**. Kinetic and thermodynamic measures of fragility in relation with coordinate of convergation point {$a;b$}.

It seems reasonable that $a$-coordinate, which represents the activation energy for "acoustic flow" by Eq.(9), i.e. when the switching bond wave (SBW) that ensures viscous flow is in a perfect accord with the acoustic bond wave (ABW) that exists also in a calm state, relates to "thermodynamic" measure, while $b$-coordinate, which describes adaptation ability of the flowing liquid by Eq.(10) relates to the "kinetic" measure of fragility.

In **Fig.15** the difference between inorganic and organic glass-formers appears most glaringly by the *opposite* pair correlations in addition to the different regions for the correlation lines. This finding agrees with the conclusion made by *Huang & McKenna* [55] who have revealed qualitatively different $\Delta C_{pg}$ *vs* $m$ correlations for IFG, OGF and polymers. Thus, different groups of glass-forming liquids, which are based on the qualitatively differing chemical bonding, should be considered separately before the general relations can be deduced. The second conclusion is that general conclusion is that "thermodynamic" and "kinetic" fragilities of a substance belonging to one group are surely related, but not more than thermodynamic and kinetic properties at all.

### 4.2. Viscous Flow under Pressure

To illustrate how convergation point works in the context of fragility, let us consider fragility as a function of pressure. This is a non-trivial task just because even the viscosity change under pressure is indefinite: viscosity may both increase (e.g., for Se [57]) and decrease (e.g., for $GeO_2$ [58]) with pressure. Fragility as such may demonstrates also a complex behavior, e.g., in the form of extrema on the $m=f(P)$ dependence at 0.5 GPa observed in OTP and salol [59].

Let us consider glycerol, whose fragility change under pressure is questionable [60]. Initial data after *Cook et al* [14] are shown in **Fig.16** in common Arrhenius coordinates.

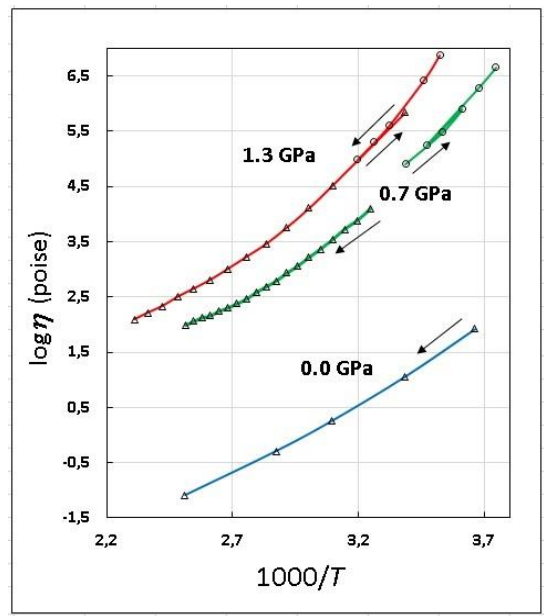

**Figure 16**. Viscosity of glycerol depending om temperature and pressure and the direction of measuring after [14].

The difference between *cooling* (arrows down) and *heating* (arrows up) regimes becomes more evident in the TA-coordinates in **Fig.17** where the 0.0 GPa line corresponds to the "normal" pressure of 1 atm (1 GPa = 10 kbar = $10^4$ atm).

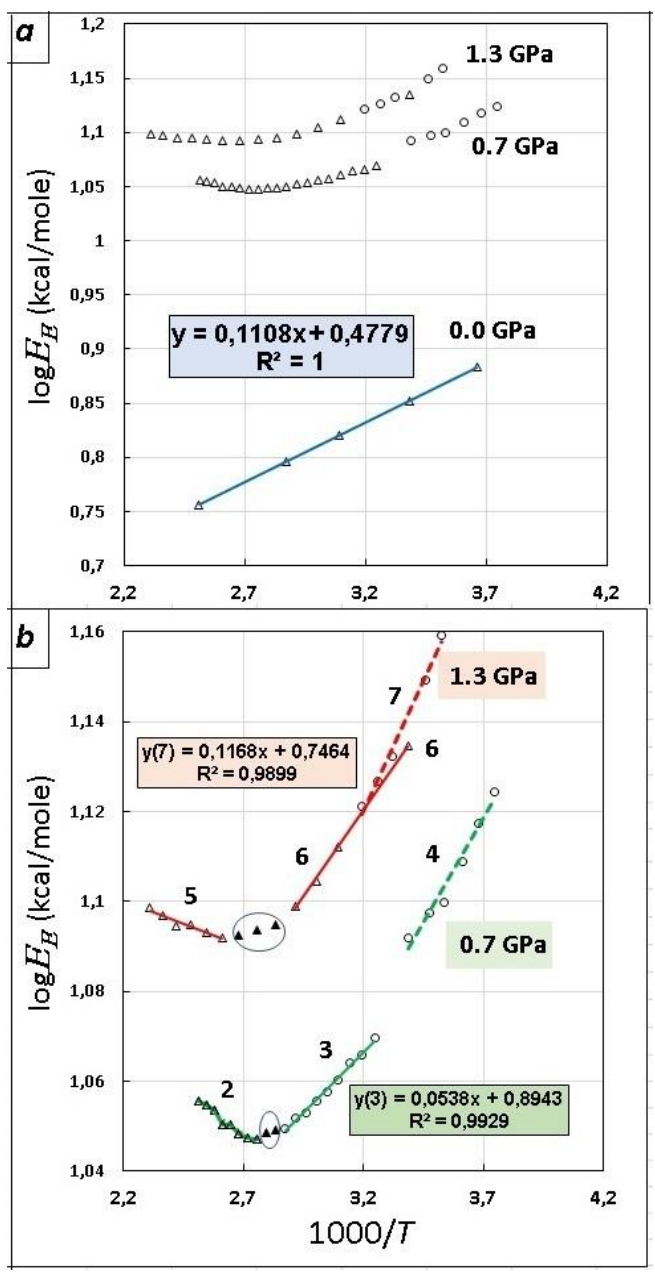

**Figure 17**. The TA-plot for glycerol under normal (*a*) and increased (*b*) pressure.

The next step is a construction of the master lines with the use of the obtained {$A_i;G_i$} points from the TA-lines equations; equations for the lines 1, 3 and 7 are shown in **Fig.17**. Since there is only one point for normal pressure (see **Fig.17a**), the addition data from other sources are used, so line 0.00 GPa in **Fig.18** repeats the master line for glycerol in **Fig.2**.

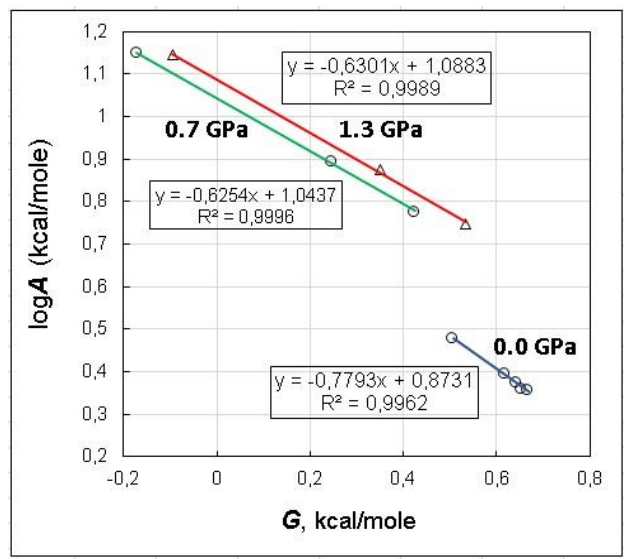

**Figure 18**. Master lines for glycerol at different pressures.

Now one can treat the viscosity-temperature-pressure data in terms of convergation point as the point shift under pressure. It is seen in **Fig.19** that the shift is large for the first jump of pressure 0.0→0.7 GPa and relatively small for the second jump 0.7→1.3 GPa, a behavior that the flowing liquid ultimately resists against pressure when the pressure becomes too high. This conclusion, however, can be made also from initial viscosity data presented in **Fig.16**. The more interesting is interpretation in terms of the bond wave model.

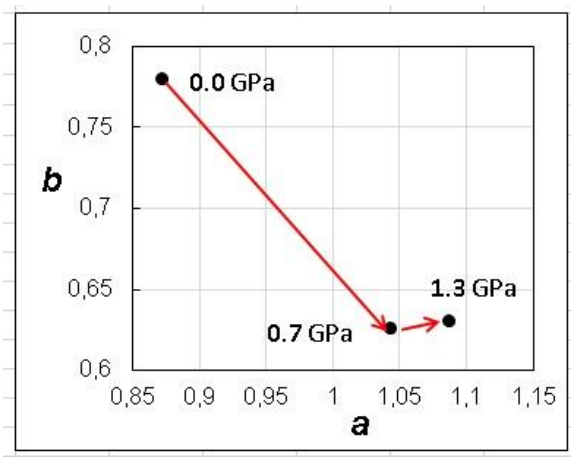

**Figure 19**. The pressure-induced shift of convergation point of glycerol.

The *a*-coordinate increases after both jumps; the increase corresponds to the growth of the acoustic bond wave energy $A^0=10^a$ from 7.5 kcal/mole at normal pressure (0.0 GPa) to 11.1 kcal mole at 0.7 GPa to 12.3 kcal/mole for final 1.3 GPa. As far as the ABW relates to the basic bonding energy, the network becomes stronger almost in two times at 1.3 GPa, but cannot increase its strength at further pressing which probably leads to a "dead" crystalline structure.

A decrease of *b*-coordinate at the first jump (0.0GPa→0.7GPa) corresponds to squeezing of the viscous space (compare "fragile" glycerol and "strong" GeO$_2$ in **Fig.3**) and decrease of adaptation ability of the flowing liquid by an appropriate change of the viscous patterns energy in accord with Eq.(10). After 0.7 GPa the *b* value remains practically constant, so the liquid stop further adaptation of viscous patterns to increasing pressure, the only one respond being a strengthening of the network in which liquid flows.

Finally, let us compare the observed shift of convergation point of glycerol with arrangement of other liquids in convergation plot. It follows from **Fig.20** that the squeezed glycerol aspires to the "strong" region, but never can reach it.

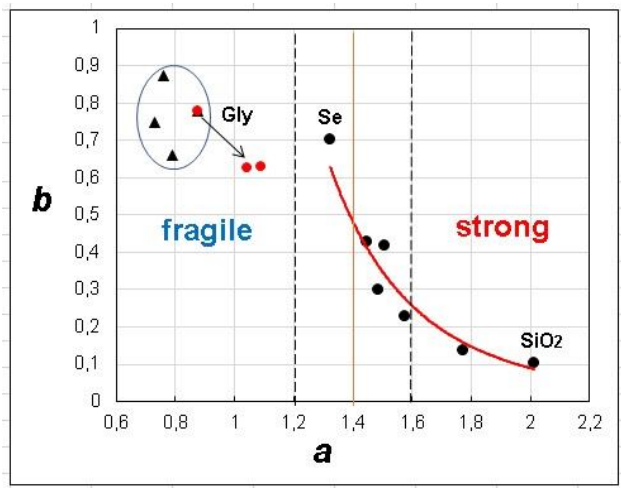

**Figure 20**. The pressure-induced shift of glycerol fragility in general convergation plot presented earlier as **Fig.4**.

The above example demonstrates how one can watch a change of both "thermodynamic" and "kinetic" fragility by means of convergation point even when the data of the heat capacity jump and the viscosity-temperature behavior near the glass transition temperature are absent.

## 5. CONCLUSION

Glass-forming liquid, being a self-organizing system, possesses an ability for a relative freedom of behavior, a feature that reveals by a partial irreproducibility of its properties. For the case of viscosity, this feature is realized by an existence of the convergation point $\{a;b\}$ that generates an infinite set of $\eta_i(T)$ relations by which a liquid adapts to concrete experimental conditions including the applied equipment and the sample history.

To obtain coordinates of convergation point for the liquid, one needs an initial set of reliable experimental data and treat them by means of the twice-activation analysis, as it is illustrated here on the example of "strong" GeO$_2$ and "fragile" glycerol including the pressure-induced shift of convergation point for glycerol. The analysis of set of typical glass-formers of different fragility, both inorganic (IGF) and organic (OGF), reveals a principal difference between the groups, which is explained from the chemical boning point of view as the difference of the basic bonds – covalent bonds for IGF and strong intermolecular bonds (hydrogen bonds and π-π interactions) for OGF – which generates alternative bonds, whose collective behavior is stipulates the bond wave. Coordinates of convergation point are related with primary acoustic and wave and the secondary switching bond wave by which glass-forming liquid flows. Finally, coordinates of convergation point are related quantitatively with "thermodynamic" and "kinetic" measures of fragilities, thus explaining the uncertainty between them and providing the "physical" sense (and also the "chemical" sense) for each.

The author hopes that this work will provide not only a new method for analysis of experimental viscosity data but also a deeper understanding of the processes in glass-formers.

## REFERENCIES

1. *C. A. Angell*. Relaxation in liquids, polymers and plastic crystals – strong/fragile patterns and problems. J. Non-Cryst. Solids 157, 131 (1991).
2. *J. C. Dyre*. Solid-that-flows picture of glass-forming liquids. J. Phys. Chem. Lett. 15, 1603 (2024).
3. *I. Avramov, A. Milchev*. Effect of disorder on diffusion and viscosity in condensed systems. J. Non-Cryst. Solids 104, 253 (1988).
4. *J. C. Mauro, Y. Yue, A. J. Ellison, et al*. Viscosity of glass-forming liquids. PNAS 106, 19780 (2009).
5. *B. N. Galimzyanov, A. V. Mokshin*. A novel view on classification of glass-forming liquids and empirical viscosity model. J. Non-Cryst. Solids 570, 121009 (2021).
6. *S. A. Dembovsky*. The connection of quasidefects with glass formation in the substances with high lone-pair electron concentration. Mater. Res. Bull. 16, 1331 (1981).
7. *S. A. Dembovsky, E. A. Chechetkina*. Glassy state clarified through chemical bonds and their defects. J. Non-Cryst. Solids 85, 346 (1986).
8. *S. A. Dembovsky, E. A. Chechetkina*. Glassy materials clarified through the eyes of hypervalent bonding configurations. J. Optoel. Adv. Mater. 3, 3 (2001).
9. *E. A. Chechetkina*. Glass clarified as the self-organizing system. arXiv 2405.00346 (2024).
10. *A. A. Mashanov, M. I. Ojovan, M. V. Darmaev, I. V. Razumovskaya*. The activation energy temperature dependence for viscous flow of chalcogenides. Appl. Sci. 14, 4319 (2024).
11. *E. A. Chechetkina*. Key distinctions in activation parameters of viscous flow for 'strong' and 'fragile' glass-forming liquids. J. Non-Cryst. Solids 201, 146 (1996).
12. *H. Eyring*. Viscosity, plasticity and diffusion as examples of absolute reaction rates. J. Chem. Phys. 4, 283 (1936).
13. *G. Tammann, W. Hesse*. The dependence of viscosity upon the temperature of supercooled liquids. Z. anorg. Allgem. Chem. 156, 245 (1926).
14. *R. L. Cook, H. E. King, Jr., C. A. Herbst, D. R. Herschbach*. Pressure and temperature dependent viscosity of two glass-forming liquids: glycerol and dibutyl phthalate. J. Chem. Phys. 100, 5178 (1994).
15. *R. Bruckner*. Physikalische Eigenschaften der oxydischen Hauptglasbildner und ihre Beziehung zur Struktur der Gläser: I. Schmelz. - und Viskositatsverhalten der Hauptglasbildner. Glastechn. Ber. 37, 413 (1964).
16. *J. P. de Neufville, C. H. Drummond III, D. Turnbull*. The effect of excess Ge on the viscosity of GeO$_2$. Phys. Chem. Glasses 11, 186 (1970).

17. *C. T. Moynihan, S. Cantor*. Viscosity and its temperature dependence in molten $BeF_2$. J. Chem. Phys. 48, 115 (1968).
18. *S. V. Nemilov, G. T. Petrovskii, G. A. Tzurikova*. Investigation of viscosity of glassy beryllium fluoride. In: Research on Chemistry of Silicates and Oxides. Leningrad, Nuaka. P.46-48 (1965).
19. *J. Malek*. Structural relaxation of $As_2Se_3$ glass by length dilatometry. J. Non-Cryst. Solids 235-237, 527 (1998).
20. *S. V. Nemilov, G. T. Petrovskii*. Investigation of viscosity of glasses of selenium-arsenic system. J. Appl. Chem. (Russ.) 36, 977 (1963).
21. *P. Kostal, B. Vcelakova, J. Malek*. Viscosity and heat capacity of $As_2S_3$ connected via Adam-Gibbs model. J. Amer. Ceram. Soc. 107, 844 (2024).
22. *O. Jantsch*. Zur Theorie der Kristallisationgeschwindigkeit unterkuhlter schmelzen. Z. Krist. 108, 185 (1956).
23. *W. T. Laughlin, D. R. Uhlmann*. Viscous flow in simple organic liquids. J. Phys. Chem. 76, 2317 (1972).
24. *M. Cukierman, W. Lane, D. R. Uhlmann*. High-temperature flow behavior of glass-forming liquids: A free-volume interpretation. J. Chem. Phys. 59, 3639 (1973).
25. *A. C. Ling, J. E. Willard*. Viscosities of some organic glasses used as trapping materials. J. Phys. Chem. 72, 1918 (1968).
26. *A. J. Barlow, J. Lamb, A. J. Matheson*. Viscous behavior of supercooled liquids. Proc. Roy. Soc. (London) A292, 322 (1966).
27. *H. Haken*. Synergetics. An Introduction (3rd ed). Springer, Berlin-Heidelberg, 1983.
28. *W. Zhang*. Selforganizology: The Science of Self-Organization. World Sci. Publ., Singapire, 2015.
29. *I. Prigogine*. From Being to Becoming: Time and Complexity in Physical Sciencies.Freeman, San Grancisco, 1980.
30. Oscillations and traveling waves in chemical systems (Eds. R. J. Field & M. Burger), Wiley, N-Y, 1985.
31. *E. A. Chechetkina*. Glassy State Clarified by the Eyes of Self-Organization. Amazon Kindle Direct Publishing, 2019.
32. *K. Stokley, M. G. Mazza, H. E. Stanley, G. Franzese*. Effect of hydrogen bond cooperativity on the behavior of water. PNAS, 107, 1301 (2010).
33. *C. A. Hunter, J. K. M. Sanders*. The nature of π-π interactions. J. Am. Chem. Soc. 112, 5525 (1990).
34. *K. E. Riley, P. Hobza*. On the importance and origin of aromatic interactions in chemistry and biodisciplines. Acc. Chem. Res. 46, 927 (2013).
35. *A. Eisenberg, A. V. Tobolsky*. Viscoelastic properties of amorphous selenium. J. Polym. Sci. 61, 483 (1962).
36. *C. Ayyagari, D. Bedrov, G. D. Smith*. Structure of atactic polysterene: A molecular dynamics simulation study. Macromol. 33, 6194 (2000).
37. *M. Kastner, D. Adler, H. Fritzsche*. Valence-alternation model for localized gap states in lone-pair semiconductors. Phys. Rev. Lett. 37, 1504 (1976).
38. *H. Haken*. Information and Self-Organization. A Macroscopic Approach to Complex Systems. Springer, Berlin-Heldenberg, 2006.
39. *H. Haken, J. Portugali*. Information and Selforganization: A Unified Approach and Applications. Entropy 18 197 (2016).
40. *D. Turnbull, M. H. Cohen*. Concerning reconstructive transformation and formation of glass. J. Chem. Phys. 29, 1049 (1958).
41. *E. A. Chechetkina*. Rawson's criterion and intermolecular interactions in glass-forming melts. J. Non-Cryst. Solids 128, 30 (1991).
42. *V. David, N. Grinberg, S. C. Moldoveanu*. In Advances in Chromatography Volume 54 (Eds.: E. Grushka, N. Grinberg), CRC Press, Boca Raton, 2018, chapter 3.
43. *B. P. Dimitrijevic, S. Z. Borozan, S. D. Stojanovic*. π-π and cation-π interactions in protein-porphyrin complex crystal structures. RSC Adv. 2, 12963 (2012).
44. *D. L. Sidebottom*. Fifty years of fragility: A view from the cheap seats. J. Non-Cryst. Solids 524, 119641 (2019).
45. *W. Oldekop*. Theoretical considerations on the viscosity of glass. Glastechn. Ber. 30, 8 (1957).
46. *C. A. Angell*. Liquid fragility and the glass transition in water and aqueous solutions. Chem. Rev. 102, 2627 (2002).
47. *E. Rossler, K.-U. Hess, V. N. Novikov*. Universal representation of viscosity in glass forming liquids. J. Non-Cryst. Solids 223, 207 (1998).
48. *V. K. Gogi, A. Mandal, A. Welton, et al*. Linking molecular origin of melt fragility index with topological phases of network glasses. J. Non-Cryst. Solids 631, 122920 (2024).
49. *C. Alba-Simionesco, G. Tarjus*. A perspective on the fragility of glass-forming liquids. J. Non-Cryst. Solids X 14, 100100 (2022).
50. *S. A. Kube, S. Sohn, R. Ojeda-Mota, et al*. Compositional dependence of the fragility in metallic glass forming liquids. Nature Comm. 13, 3708 (2022).
51. *R. Bohmer, K. L. Ngai, C. A. Angell, D. J. Plazek*. Nonexponential relaxations in strong and fragile glass formers. J. Chem. Phys. 99, 4201 (1993).
52. *S. S. N. Murthy*. Strength and fragility in glass-forming liquids. J. Phys. Chem. 93, 3347 (1989).
53. *C. A. Angell, W. Sichina*. Thermodynamics of the glass transition: Empirical aspects. Ann. NY Acad. Sci. 279, 53 (1976).
54. *D. L. Sidebottom*. Connecting glass-forming fragility to network topology. Front. Mater. 00144 (2019).
55. *D. Huang, G. B. McKenna*. New insights into the fragility dilemma in liquids. J. Chem. Phys. 114, 5621 (2001).
56. *U. Senapati, A. K. Varshneya*. Viscosity of chalcogenide glass-forming liquids: An anomaly in the 'strong' and 'fragile' classification. J. Non-Cryst. Solids 197, 210 (1996).
57. *D. E. Harrison*. Effect of pressure (up to 4 kbar) on the polymerization of liquid selenium from measurements of viscosity. J. Chem. Phys. 41, 844 (1964).
58. *S. K. Sharma, D. Virgo, I. Kushiro*. Relationship between density, viscosity and structure of $GeO_2$ melts at low and high pressures. J. Non-Cryst. Solids 33, 235 (1979).
59. *K. U. Schug, H. E. King, Jr, R. Bohmer*. Fragility under pressure: Diamond anvil cell viscometry of ortho-terphenyl and salol. J. Chem. Phys. 109, 1472 (1998).
60. *S. Pawlus, M. Paluch, J. Ziolo, C. M. Roland*. On the pressure dependence of the fragility of glycerol. J. Phys.: Condens. Matt. 21, 332101 (2009).